# Broadband dielectric spectroscopy on single-crystalline and ceramic CaCu$_3$Ti$_4$O$_{12}$

S. Krohns, P. Lunkenheimer[a]
*Experimental Physics V, Center for Electronic Correlations and Magnetism, University of Augsburg, 86135 Augsburg, Germany*

S. G. Ebbinghaus
*Solid State Chemistry, University of Augsburg, 86135 Augsburg, Germany*

A. Loidl
*Experimental Physics V, Center for Electronic Correlations and Magnetism, University of Augsburg, 86135 Augsburg, Germany*

We present dielectric measurements of the colossal dielectric constant material CaCu$_3$Ti$_4$O$_{12}$ extending up to 1.3 GHz also covering so far only rarely investigated single crystalline samples. Special emphasis is put on the second relaxation reported in several works on polycrystals, which we detect also in single crystals. For polycrystalline samples we provide a recipe to achieve values of the dielectric constant as high as in single crystals.

Among the vast number of papers on the extremely high ("colossal") dielectric constants ($\varepsilon'$) found in CaCu$_3$Ti$_4$O$_{12}$ (CCTO) there are at least ten so-called "highly-cited" papers[1,2,3,4,5,6,7,8,9,10]. This demonstrates the tremendous interest in new high-$\varepsilon'$ materials, which are prerequisite for further advances in the development of capacitive electronic elements. It soon became clear that the colossal $\varepsilon'$ in CCTO must have a non-intrinsic origin[4,5,6,7,8,9,11,12,13]. Nowadays usually the results are interpreted within an "internal barrier layer capacitor" (IBLC) picture: Polarization effects at insulating grain boundaries between semiconducting grains or other internal barriers generate non-intrinsic colossal values of $\varepsilon'$, accompanied by a strong Maxwell-Wagner (MW) relaxation mode. As an alternative, a "surface barrier layer capacitor" (SBLC) picture was proposed, assuming, e.g., the formation of Schottky diodes at the contact-bulk interfaces[8,9,14].

Nearly all experimental evidence for a non-intrinsic mechanism so far is based on measurements of ceramic samples (e.g., refs. 4,5,6,7,9,11,12,13). However, extremely high values of $\varepsilon'$ of the order of 10$^5$ were observed particularly in CCTO single crystals (SCs)[3]. With only few exceptions[11], measurements of polycrystals (PCs) reveal much lower values in the range of 10$^3$ – 10$^4$. There are no grain boundaries in SCs and thus there were speculations about other internal boundaries as, e.g., twin-boundaries[4,6,15,16]. However, if there is any kind of planar defects in SCs generating strong relaxations, should they not contribute to a separate relaxational response in PCs, where grains can reach sizes up to 100 µm, too? Interestingly, it seems clear now that indeed there is a second relaxation in CCTO PCs[9,12,14] and it even was already observed in one of the earliest reports on CCTO[2]. It leads to even larger $\varepsilon'$ values than the well-known main relaxation. Thus one could speculate that one of the two relaxations is due to grain boundaries and the other due to planar defects within the grains. Alternatively, one relaxation could be due to an IBLC and the other to a SBLC mechanism[9]. In the present work we aim to elucidate differences and similarities in SC and PC behavior, address the question of the second relaxation, and try to help solving the IBLC/SBLC controversy. For this purpose, we performed measurements on various samples including single-crystals, which to our knowledge so far only were investigated in two works[3,16]. The spectra cover up to nine frequency decades and extend up to the technically relevant GHz frequency range.

Polycrystalline samples of CCTO were prepared as reported in ref. 9 and sintered at 1000°C in air for up to 48 h. SCs were grown by the floating zone technique[3]. The applied growth furnace is equipped with two 1000 W halogen lamps, the radiation of which is focused by gold-coated ellipsoidal mirrors. Polycrystalline bars serving as seed and feed rods were cold-pressed and sintered in air for 12 h at 1000 °C. The seed rod was rotated with a speed of 30 rpm, while the feed was kept still. The growth rate was adjusted to 5 mm/h. To avoid thermal reduction of copper, crystal growth was performed in oxygen (flow rate 0.2 l/min) at a pressure of 4 bar. For the dielectric measurements silver paint or sputtered gold contacts (thickness 100 nm) were applied to adjacent sides of the platelike samples. The complex conductivity and permittivity were measured over a frequency range of up to nine decades (1 Hz < $\nu$ < 1.3 GHz) at temperatures down to 40 K as detailed in refs. 9,17.

Fig. 1 shows broadband dielectric results for single crystalline CCTO with sputtered contacts. Colossal values of $\varepsilon' > 10^5$ are achieved. In $\varepsilon'(\nu)$, the well-known[2,3,4,9] strong relaxation steps are observed. Correspondingly, peaks in the loss $\varepsilon''(\nu)$ and shoulders in the conductivity $\sigma'(\nu)$ show up. These features can be explained by a MW relaxation caused by an equivalent circuit consisting of the bulk contribution, connected in series to a parallel RC circuit. At high frequencies, the highly resistive thin layers generating this RC circuit (planar defects or surface layers[16]) are shorted via their capacitance and intrinsic behavior is detected. It reveals an $\varepsilon'$ of the order of 100 and a succession of dc and ac conductivity, characteristic for hopping of localized charge carriers[9,18]. Our results demonstrate that as for PCs[9], also for SCs the room temperature value of $\varepsilon'$ is dramatically reduced at high frequencies, which represents a problem for possible high-frequency applications of CCTO.

---

[a] Electronic mail: peter.lunkenheimer@physik.uni-augsburg.de

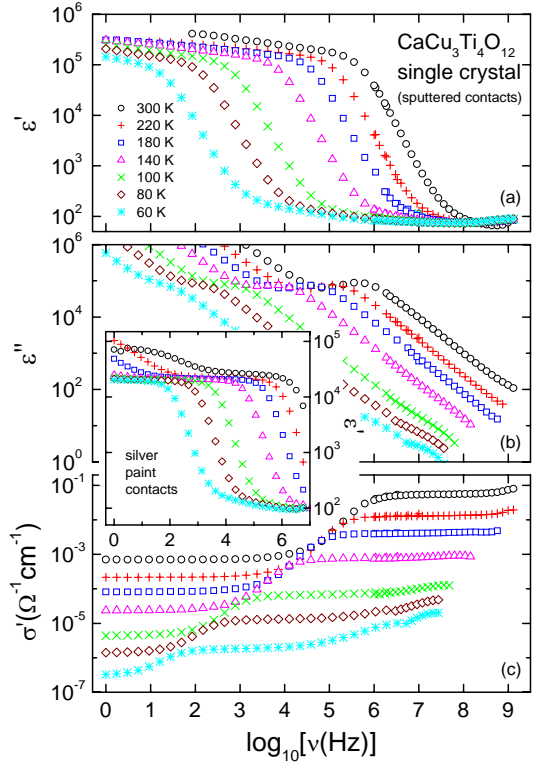

FIG. 1. Frequency-dependent dielectric constant (a), loss (b), and conductivity (c) of single-crystalline CCTO with sputtered gold contacts at various temperatures. The inset shows $\varepsilon'$ for the same sample using silver paint contacts.

The inset of Fig. 1 demonstrates that as for PCs[9], also in SCs silver paint contacts lead to much lower upper plateau values of the relaxation mode ($\varepsilon'_{high} \approx 2\times10^4$ in the present case). In PCs this was interpreted assuming a surface-related origin of the high $\varepsilon'$ values[9]. One may argue that due to their grain structure, silver-paint contacts are not suited for the measurements of very high dielectric constant materials. However, we demonstrated[9,19] that for intrinsic $\varepsilon'$ values up to $2\times10^4$ and $2\times10^5$, respectively, silver paint contacts lead to the same results as sputtered gold. Another very interesting finding revealed by the inset of Fig. 1 is the occurrence of a second relaxation, which to our knowledge never before was observed in SCs. Indications for a second relaxation also show up in Fig. 1(a) for sputtered contacts, but the related very high capacitance values partly are out of the resolution window of the measuring device. The occurrence of a second relaxation also in SCs excludes the above-mentioned scenario of one of the two relaxations being due to grain boundaries (present in PCs only) and one being due to internal boundaries (present in SCs and PCs).

Now the question arises how to achieve $\varepsilon'_{high}$ values in ceramic CCTO, comparable to those in single-crystalline samples. In light of the SBLC mechanism, assuming diode formation at the surface, sputtered contacts should be used[8,9]. Within the IBLC framework, the grain size should be increased by suitable tempering procedures, as reported in refs. 11,12,13. We combined both procedures, the results being shown in Fig. 2. Overall, a behavior quite similar to that of the SC (Fig. 1) is obtained, especially concerning the absolute values of $\varepsilon'_{high}$ of the order of $10^5$. They belong to the highest ever observed in ceramic CCTO[11]. Also the intrinsic behavior is quite similar, namely the intrinsic $\varepsilon'$ (somewhat below 100), the dc conductivity (about 4 - $5\times10^{-2}$ $\Omega^{-1}$cm$^{-1}$ at 300 K), and the ac conductivity, $\sigma' \sim \nu^s$ ($s<1$) due to hopping transport. There are some minor differences concerning the MW relaxation, e.g., the width of the relaxation features is larger in the PC, most likely due to the less homogeneous material leading to a distribution of relaxation times.

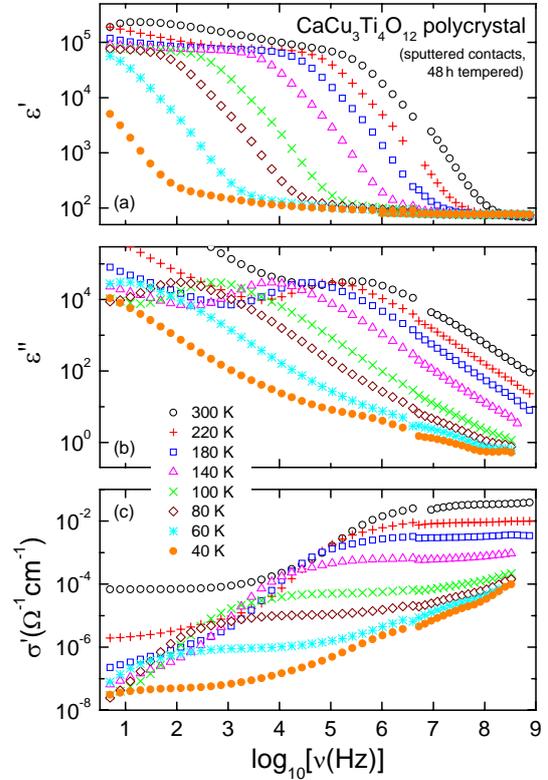

FIG. 2. $\varepsilon'(\nu)$ (a), $\varepsilon''(\nu)$ (b), and $\sigma'(\nu)$ (c) of polycrystalline CCTO (tempered 48 h) with sputtered gold contacts at various temperatures.

Also in the results of Fig. 2, the second relaxation shows up. We performed a variety of experiments on PCs to check the influence of different sample treatments on this relaxation. We found a very strong effect when removing the outer surface layers of a PC[20], reducing the original sample thickness of 1.63 mm by about 0.08 mm. The polishing was done under $N_2$ atmosphere to avoid any reaction with oxygen at the freshly exposed surface. Afterwards the sample was installed in the vacuum of the cryostat as quickly as possible. Fig. 3 shows the results for the untreated sample (a) and after polishing (b). The weak second relaxation, barely visible in the untreated sample, becomes very prominent after removal of the surface layer. After the experiment of Fig. 3(b), the sample was exposed to air for 48 h. In the subsequent measurement [Fig. 3(c)] the amplitude of the second relaxation seems to have become smaller again and a strong shift of the relaxation frequencies is observed. Also the conductivity varies considerably for the different sample treatments of Fig. 3. This is revealed by the inset, which shows as an example $\sigma'(\nu)$ at 140 K for the three different states of the PC. Only at frequencies $\nu > 10^5$ Hz, where intrinsic behavior dominates, the curves agree. Overall, the results of Fig. 3 indicate that surface effects contribute in some way to the generation of the second relaxation of CCTO. One could speculate, e.g., about an insulating surface layer due to a deviating stoichiometry, which is modified by exchange with air and was partly or completely removed during polishing. Further systematic experiments are necessary to clarify this issue.

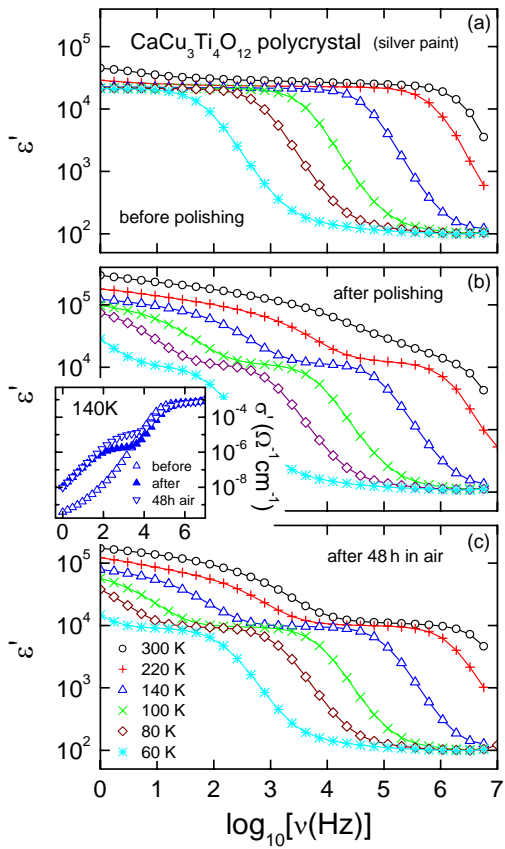

FIG. 3. $\varepsilon'(\nu)$ of a ceramic CCTO sample with silver paint contacts at various temperatures. Results are shown for the untreated sample (a), after polishing under $N_2$ atmosphere (b), and after leaving it in air for 48 h (contacts removed) (c). The inset shows the conductivity at 140 K for the three treatments.

In summary, we performed dielectric measurements up to 1.3 GHz on various CCTO samples also comprising so far only rarely investigated single-crystalline samples. In SCs we detected a reduction of $\varepsilon'$ at MHz-GHz frequencies, the occurrence of a second relaxation, and an enhancement of $\varepsilon'$ for sputtered contacts, features that so far were observed in polycrystalline CCTO only. We provided a strategy for the enhancement of $\varepsilon'$ in PCs to values comparable to those of SCs. The occurrence of both relaxational features also in SCs leads to the conclusion that none of them can be exclusively ascribed to grain boundaries. Instead, planar crystal defects (e.g., twin boundaries), surface contributions (e.g., Schottky diodes), or a combination of both must be considered. Our contact- and polishing-dependent measurements in SCs and PCs seem to indicate that surface-effects at least play some role. It is clear that our results are not in accord with the numerous experiments providing evidence for an IBLC mechanism that is due to grain-boundaries in CCTO, e.g. the variation of $\varepsilon'$ with grain size[11,12,13] or the microcontact measurements within single grains and across individual grain boundaries in PCs[7]. At the present stage, we have no solution for these discrepancies but we believe it would be unreasonable to assume different mechanisms for SCs and PCs. Obviously, further experiments, especially at SCs, are necessary to achieve a thorough understanding of the dielectric properties of CCTO.

This work was supported by the Commission of the European Communities, STREP: NUOTO, NMP3-CT-2006-032644 and by the DFG via the SFB 484.